# ESR and TSL study of hole capture in PbWO$_4$:Mo,La and PbWO$_4$:Mo,Y scintillator crystals

V.V. Laguta[1], M. Nikl[1], J. Rosa[1], D. Savchenko[1], S. Zazubovich[2]

[1]Institute of Physics AS CR, Cukrovarnicka 10, 162 53 Prague, Czech Republic
[2]Institute of Physics, University of Tartu, Riia 142, 51014 Tartu, Estonia

**Abstract**

The processes of hole localization in double-doped PbWO$_4$:Mo,La and PbWO$_4$:Mo,Y single crystals have been studied by continuous wave and pulse electron spin resonance (ESR) and thermally stimulated luminescence (TSL) methods. We show that the holes created by the UV irradiation are preferably trapped at lattice oxygen ions in the vicinity of perturbing defects such as lead vacancies, impurity ions (La, Y, Mo), and other lattice imperfections. This leads to a variety of O$^-$ centers, which differ both by thermal stability (from about 170 K up to 240 K) and ESR parameters. The hole centers of this type were not observed neither in PbWO$_4$:Mo nor in PbWO$_4$:La(Y) crystals. The recombination processes of thermally released holes with electrons stored at different traps, including Pb$^+$-WO$_3$ and (MoO$_4$)$^{3-}$ centers, are systematically studied by TSL. Thermal stability parameters are defined by ESR and TSL techniques for different O$^-$ type defects.

# I. INTRODUCTION

PbWO$_4$ (PWO) is a high-density scintillator with a fast intrinsic luminescence, which is successfully applied in high-energy physics detectors. In this material, the intrinsic luminescence is of excitonic nature, and charge transfer transitions at the (WO$_4$)$^{2-}$ oxyanion complex give rise to the emission in the blue spectral region [1-3]. It was shown in [4] that the blue emission spectrum consists of strongly overlapping bands of the self-trapped exciton and various localized excitons. Mechanism of creation of the localized excitons and their photo-thermally stimulated disintegration in PWO has been recently reported in great details [4-5].

Because of the mentioned disintegration of the exciton-based emission centers, any shallow trap states in the PWO lattice, taking part in carrier capture processes, become very important. They modify the migration characteristics of free charge carriers through re-trapping. Consequently, the speed of their delayed radiative recombination is altered. The simultaneous use of TSL and ESR experiments has proved efficient to gain understanding about the nature of traps and details about the energy storage in the PWO lattice. Specifically, electron traps are well investigated in this crystal (see, e.g., the last review [6]). However, the question about the origin of the recombination partners in the TSL processes, i.e., the counterpart hole traps, remains weakly understood.

Due to favorable properties of PWO (a high density, fast response, low cost), some research laboratories tried to increase the light yield of PWO by doping [7-8]. Double doping by Mo$^{6+}$ and A$^{3+}$ (A=La, Y) ions appeared as the most successful way, providing a light yield increase by a factor of 2–3 [8-9] without deterioration of other scintillation characteristics. Similarly to undoped PWO, a favorable influence of a trivalent rare-earth ion co-doping was found, evidenced by a lower concentration of trapping states and higher radiation hardness with respect to PWO:Mo [9-10]. However, the defect compensation appeared less efficient than in the undoped material, and a Mo clustering in PWO:Mo was noticed at high Mo concentrations [9, 11]. Hence, the defects nature and their role in the energy transfer and storage in PWO:Mo,A$^{3+}$ deserves further studies. The origin of a composite TSL glow peak observed in [5, 12] in the PWO:1200 ppm Mo, 80 ppm La crystal in the 170–250 K range requires further investigations. In previous studies carried out for different undoped and Mo-doped PWO crystals, all the TSL peaks in this temperature range were ascribed to the thermal destruction of various electron centers, e.g., Pb$^+$-WO$_3$ centers [13], oxygen vacancy clusters with two trapped electrons [14-16], (MoO$_4$)$^{3-}$ centers [12, 15], etc. In these TSL peaks, the green G(II) emission (2.5 eV, FWHM=0.55 eV [17]) is observed ascribed in [18] to the oxygen-deficient anion complexes in the form of WO$_3$. As it has been shown in [19, 20], this emission accompanies the optically and thermally stimulated tunneling recombination of electron and hole centers in the 150-300 K temperature range. Thus, the origin of the G(II) emission is strongly different from that of the relatively fast green G(I) emission observed at T<200 K under photoexcitation in the Mo-related absorption band [17].

An indirect indication about existence of hole defects has been obtained in Mo,La-doped PWO crystals [12]. There are also indirect ESR indications on the creation of hole centers in other PWO crystals. For example, in the crystals grown by the Bridgman method, the thermal destruction of the hole centers at 200-220 K is reflected in the lowering concentration of $(MoO_4)^{3-}$ and $(CrO_4)^{3-}$ electron centers [14] which are thermally stable up to much higher temperatures. Due to the low probability of the hole self-trapping in the PWO lattice [21], a hole stabilization by a defect seems to be necessary. Possible non-paramagnetic character of the hole states suggests that a couple of trapped holes could be stabilized around a lead vacancy ($V_{Pb}$) to restore the local charge balance [22, 23]. Such nonparamagnetic ($V_{Pb}$+2h)-type hole centers can be responsible for some TSL peaks detected at 220-230 K [24].

Recently, we provided a direct evidence of an existence of intrinsic (related to regular lattice ions) hole centers in PWO doped with Mo and La or Y by observation of their ESR spectra [25]. It was found that holes created by the UV irradiation are trapped at a lattice oxygen ion nearby a lead ion vacancy ($V_{Pb}$) and/or an impurity ion (e.g., La, Y, Mo), forming a variety of $O^-$ centers with thermal stability from about 170 K up to 240 K. This enables the radiative recombination of freed holes with the electrons localized at different traps, including $Pb^+$-$WO_3$ and $(MoO_4)^{3-}$ centers, accompanied with the appearance of the TSL peaks located around 190 K and in the 225-250 K temperature range. In the present paper, characteristics of the $O^-$-related hole centers in PWO:Mo,La and PWO:Mo,Y crystals are studied in detail by using correlated TSL and ESR measurements on the same crystals. In particular, we have determined all spectral characteristics of $O^-$ centers. From measurement of concentration decay of trapped holes we have determined thermal stability parameters such as ionization energies and frequency factors. These parameters are in fair agreement with similar data obtained from analysis of TSL peaks.

**II. EXPERIMENTALS**

Two groups of Mo,La-doped PWO crystals were used in our investigation. The first one, #1, was grown in Furkawa Co., Ltd., Japan (≈1200 ppm of Mo and 80 ppm of La in the crystal) from high purity (5N) raw materials by the Czochralski method in air, using a platinum crucible and the third crystallization method [26], and previously studied in [5, 12]. The second group of crystals, #2, (≈40 ppm of Mo and 120 ppm of La) was grown in CRYTUR Ltd., Czech Republic, from 5N purity raw materials by the same Czochralski method. The PWO:Mo,Y crystal, containing 1000 ppm Mo and 100 ppm Y, and PWO:Mo crystal, containing 1000 ppm Mo were grown in Institute of Physics, Chech Republic by the Czochralski method as well. For comparison, some measurements were made at the undoped $PbWO_4$ crystal grown in Furukawa Co. and studied in [13, 24].

The PbWO$_4$ crystal structure is characterized by the tetragonal space group I4$_1$/a or $C_{4h}^6$ with the unit-cell dimensions $a = b = 5.456$ Å and $c = 12.020$ Å. The W and Pb sites have the S$_4$ point symmetry. The O sites have only the trivial point symmetry. They are arranged in the tetrahedral coordination around each W site. The WO$_4$ tetrahedra are rotated around the $c$ axis by an angle of 14$^0$. The main crystallographic parameters of PWO can be found, e.g., in Ref. [27].

For ESR measurements, the crystals were oriented, cut in the (100) and (001) planes in a typical shape of about 2.5×2.5×8 mm$^3$, and polished. The continuous wave (CW) ESR measurements were performed at 9.25-9.8 GHz with the standard 3 cm wavelength of the ESR spectrometer in the temperature range 10-290 K using an Oxford Instrument cryostat. The pulsed ESR experiments were performed with a Bruker E580 spectrometer with a dielectric resonator. A mercury high-pressure arc lamp with optical filters was used for the UV irradiation of the samples.

For the thermally stimulated luminescence study, the crystal was selectively irradiated at 140-230 K in the 4.7-3.5 eV energy range with the deuterium DDS-400 lamp through a monochromator (the spectral width of the slit was 5 nm). The TSL curves were measured in the 140-300 K temperature range with the heating rate of 0.2 K/s$^{-1}$. The TSL intensity was detected by a photomultiplier with an amplifier and recorder.

### III. EXPERIMENTAL RESULTS AND DISCUSSION
#### A. CW ESR spectra

As is has been shown in our previous publication [25] the broadband UV irradiation of both PWO:Mo,La and PWO:Mo,Y with wavelength of 270-370 nm at T≈165 K induces, along with (MoO$_4$)$^{3-}$ [12], a paramagnetic defect which was assigned to the O$^-$ hole centers. When the irradiated crystal is heated up to the temperature about 185 K, this spectrum disappears and other spectra appear. Under gradual heating from 165 to 240 K, at least seven types of such spectra can be distinguished in PWO:Mo,La #1 sample while in other PWO:Mo,La and PWO:Mo,Y crystals only two types of such O$^-$ defects, O$_I$ and O$_{II}$, can be created at detectable concentration. For distinctness, we have denoted all visible hole centers as O$_I$ up to O$_{VII}$. Note that the concentration of the created centers strongly depends on the irradiation temperature, so that O$^-$ centers, like (MoO$_4$)$^{3-}$ centers [12], can be effectively created only in the narrow temperature range (see, e.g., Fig. 1).

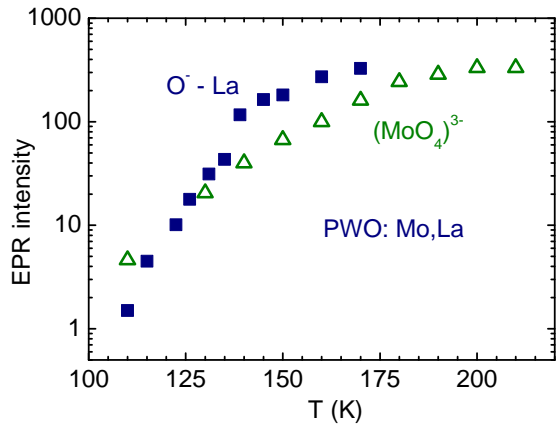

Fig. 1. Dependence of the O$^-$ (O$_I^{La}$) and (MoO$_4$)$^{3-}$ ESR intensity on the irradiation temperature.

The spectral characteristics of the hole centers as determined from the fit of experimental angular dependencies of the resonance fields are presented in Table I. The g tensor components, excepting the $O_{VII}$ center, are typical for the $O^-$-type hole centers [28] and are close in value to those reported for isostructural $CaWO_4$ crystals where holes, however, are self-trapped [29].

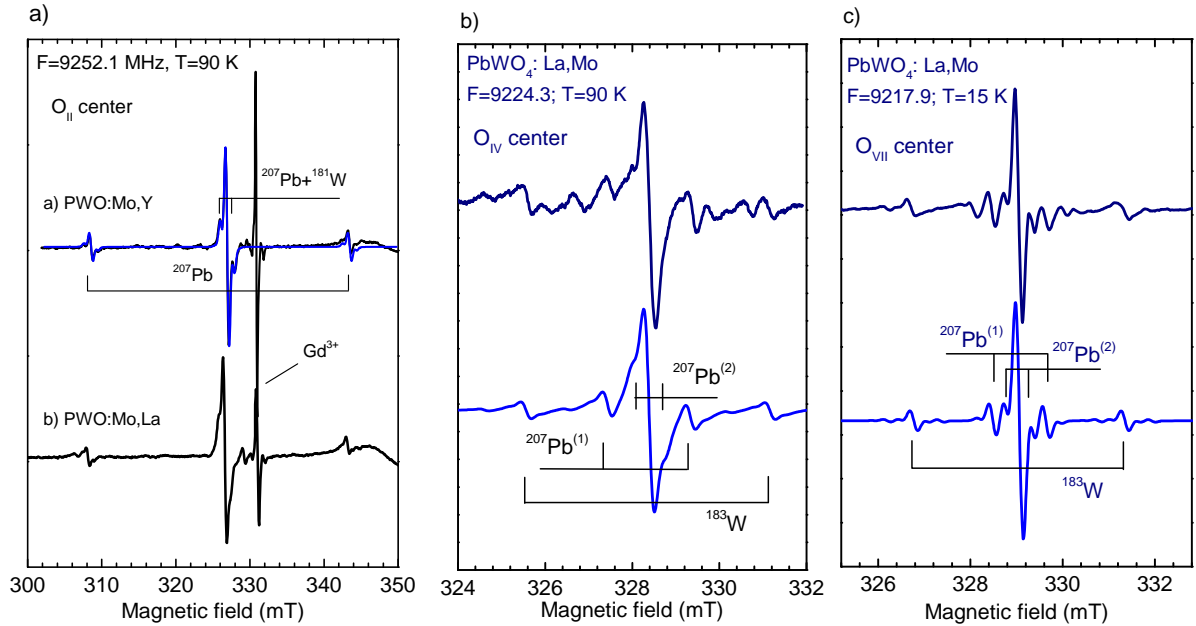

Fig. 2. Measured (points) and simulated (solid lines) ESR spectra of $O_{II}$ (a), $O_{IV}$ (b), and $O_{VII}$ (c) hole centers showing the hf structure from $^{183}W$ and $^{207}Pb$ isotopes. A bit different structure of the central components of $O_{II}$ center in PWO:Mo,La and PWO:Mo,Y is result of small deviation of crystal orientation **B** II **c**, where four spectral lines merge into one.

In Table 1, we have also included the hyperfine (hf) constants in the cases when the corresponding hf splitting due to the nearest nuclei with nonzero nuclear magnetic moment could be resolved and analyzed with the sufficient reliability. As an example, Fig. 2 shows the simulation of the hf structure for ESR spectra of $O_{II}$ center in PWO:Mo,La and PWO:Mo,Y crystals and $O_{IV}$ and $O_{VII}$ centers in PWO:Mo,La #1 crystal. The electron-nuclear interactions being nearly isotropic are in the range from 0.5 mT for $^{207}Pb$ up to 5.6 mT for $^{183}W$ nuclei. However, much stronger (~35 mT) hf splitting was revealed in the case of $O_{II}$ center (see Fig. 2(a)). These hf lines were overlooked in our previous study [25] due to their really huge splitting. The intensities of these hf lines correspond to the interaction of a hole with $^{207}Pb$ nucleus. As the $O_{II}$ spectrum is practically the same in the both types of PWO crystals, doped with La or Y, one can conclude that it is produced by hole centers of the same origin. Complete simulation of the hf structure (Fig. 2(a)) indicates that hole effectively interacts at least with two $^{207}Pb$ and one $^{181}W$ nuclei. The local structure of this center can be similar to the $V_k$ center in $CaWO_4$ [29]. However, essential difference from the $V_k$ center in $CaWO_4$ is that the trapped hole is mainly delocalized within only one tetrahedron. The electron deficiency is also shifted from oxygen ion to the closest lead ion producing high hf field at Pb nucleus. The second neighboring tetrahedron can be occupied by a Mo impurity. Such perturbing defect is necessary in order to stabilize hole at a local energy level. Although $^{95,97}$Mo hf lines were not resolved in the CW spectrum due to weak their intensity and probably small hf splitting comparable with the linewidth, the presence of Mo in the vicinity of O$^-$ ion can not be excluded as the crystal was doped by this impurity. 2D HYSCORE spectrum shows that even $O_I^Y$ center contains Mo ion at neighborhood of trapped hole (see section B).

### B. Pulse EPR spectra and hyperfine sublevel correlation spectroscopy of $O_I$ hole center

In contrast to the $O_{II}$ center, which seems to have similar structure in both the La and Y doped PWO crystals, the structure of the $O_I$ center essentially depends on type of impurity doping. In particular, hf structure from $^{139}$La isotope is well resolved in La doped crystals [25]. However, expected hf structure from $^{89}$Y nuclei is not resolved in Y doped crystals. In previous publication [25] this fact was reasonably explained by a small value of $^{89}$Y nuclear magnetic moment so that the expected $^{89}$Y hf splitting is smaller than the linewidth. In order to confirm our assignment of the $O_I$ spectrum to $[Y^+_{Pb}+V^{2-}_{Pb}]+O^-$ defect, the PbWO$_4$:Mo,Y crystals were investigated by echo-detected electron paramagnetic resonance (ED EPR) and hyperfine sublevel correlation spectroscopy (HYSCORE) pulse techniques.

It is well known that HYSCORE is a useful tool to resolve small hf couplings of distant nuclei or nuclei with weak magnetic moments that cannot be resolved in the ordinary EPR spectra [30]. The four-pulse sequence $\pi/2-\tau-\pi/2-t_1-\pi-t_2-\pi/2-\tau$–echo used in this technique correlates the nuclear magnetic resonance transitions $\nu_{\alpha i}$ within one electron spin manifold $M_\alpha$ with the nuclear transitions $\nu_{\beta i}$ of the other electron spin manifold $M_\beta$. In case of isotropic or nearly isotropic hf interaction, these correlations appear in the 2D spectrum as off-diagonal cross peaks with frequencies at ($[\nu_\alpha,\nu_\beta]$; $[\nu_\beta,\nu_\alpha]$) in the weak coupling case (A<2$\nu_I$, $\nu_I$ is the Larmor frequency of a nuclear spin) and at ($[-\nu_\alpha,\nu_\beta]$; $[-\nu_\beta,\nu_\alpha]$) for the strong-coupling case (A>2$\nu_I$), which are symmetric with respect to the diagonal and antidiagonal lines, respectively (see Fig. 3).

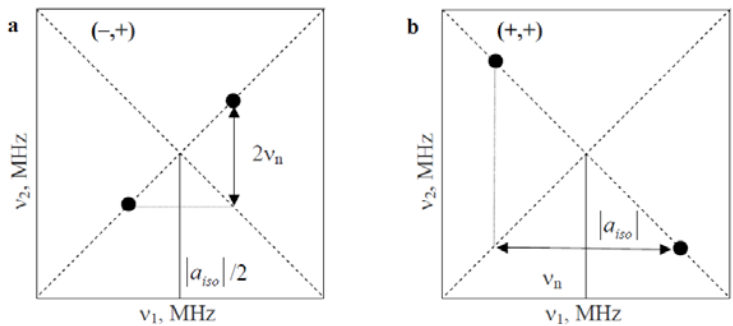

Fig. 3. Schematic representation of the HYSCORE patterns cross-peaks for isotropic hf coupling and S = 1/2, I = 1/2 for (a) strong-coupling case (A>2$\nu_I$) and

(b) weak-coupling case ($A < 2\nu_I$).

For the system with $S = 1/2$ and $I > 1/2$, the double-quantum transitions appear at frequencies [31]:

$$\nu_{dq} = 2[(\nu_I \pm |A|/2)^2 + P^2(3+\eta^2)]^{1/2}, \quad (1)$$

where $P = e^2qQ/(4\hbar)$ is the nuclear quadrupole coupling frequency, and $\eta$ is the asymmetry parameter of the nuclear quadrupole interaction. Often, only these double-quantum transitions are well visible while the single-quantum transitions are practically unresolved.

Finally, the dipole electron-nuclear hyperfine parameter B for system with axially symmetric **g** tensor can be calculated from the maximum vertical shift $\Delta\nu_s$ of the cross peaks from the antidiagonal line [32]:

$$B = \frac{2}{3}\sqrt{\frac{8\Delta\nu_s\nu_I}{\sqrt{2}}}. \quad (2)$$

The ED EPR spectra were measured using two-pulse Hahn echo sequence: $\pi/2 - \tau - \pi - \tau$ –echo with the pulse lengths: $\pi/2$ = 16 ns, $\tau$ = 400 ns, $\pi$ = 32 ns. The pulse lengths for HYSCORE experiment were: $\pi/2$ = 16 ns, $\pi$ = 28 ns, $t_1 = t_2$ = 400 ns, time increment 30 ns. The $\tau$ values were: 160, 500, 680 and 800 ns. The spectra resolution was [256x256] points.

The ED EPR spectrum measured in PbWO$_4$:Mo,Y crystal at T = 18 K after the UV light irradiation at T = 155 K and further cooling down of the sample at dark is shown in Fig. 4. Note that for maximum intensity of the ED spectrum from the O$^-$ ions, the measurements were performed at B II [001], where four lines from four crystallographically equivalent centers merge into one spectral line. The ED spectrum contain contributions from three paramagnetic centers: O$_I$, Gd$^{3+}$ and (MoO$_4$)$^{3-}$, in agreement with CW measurements.

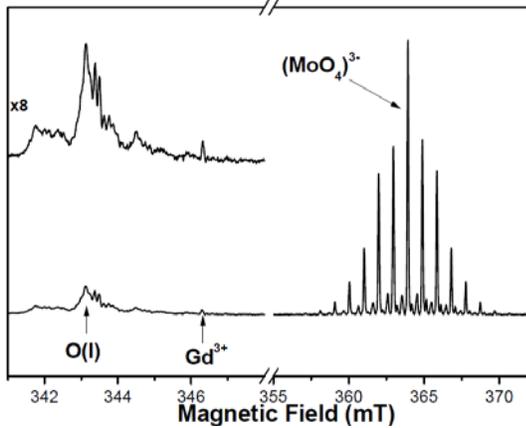

Fig. 4. The X-band ED EPR spectrum measured in PbWO$_4$:Mo,Y crystal at B II [001]) and T = 18 K after irradiation of the crystal with UV light at T = 155 K. O$_I$ spectrum is slightly distorted due to contribution of O$_{VII}$ center.

The sum of HYSCORE spectra measured at the ED EPR line position of O$_I$ center (B$_0$ = 343.12 mT) with four different $\tau$ values is presented in Fig. 5. It is seen from this figure that there are three cross-peaks in (–, +) quadrant (strong-coupling case). The intense cross-peaks at ([–6.71, 1.82], [–1.82, 6.71]) according to Eq. (1) were tentatively attributed to the double-quantum transition of $^{95}$Mo (I = 5/2, 15.92 %, $\nu_I$ = 0.96 MHz) and $^{97}$Mo (I = 5/2, 9.55%, $\nu_I$ = 0.98 MHz) nuclei with small quadrupole contribution. The cross-peaks at ([–4.30, 2.54], [–2.54, 4.30]) were attributed to hf interaction of hole with $^{89}$Y (I = 1/2, 100 %, $\nu_I$ = 0.72 MHz) nuclei and the cross-peaks at ([–1.82, 0.52], [–0.52, 1.82]) can be assigned to hf interaction with $^{183}$W (I = 1/2, 14.31%, $\nu_I$ = 0.62 MHz) nuclei.

In the (+, +) quadrant (weak-coupling case), there are six cross-peaks at ([0.65, 6.12], [6.12, 0.52]); ([0.65, 5.47], [5.47, 0.65]); ([0.98, 5.34], [5.34, 0.98]); ([1.30, 5.27], [5.27, 1.30]); ([1.69, 4.43], [4.43, 1.69]) and ([2.60, 3.65], [3.65, 2.60]) which are clearly related to hf interaction of a hole with $^{207}$Pb (I = 1/2, 22.1%, $\nu_I$ = 3.07 MHz) nuclei. Some of $^{207}$Pb peaks are vertically shifted from the $\nu_1 = -\nu_2$ axis. This shift is due to the dipole hf interaction.

The hf parameters for $O_I$ center with the surrounding nuclei in the [001] orientation obtained from HYSCORE spectra analysis are represented in Table II. The B values were roughly estimated from the cross peaks shift. It should be also noted that the spectral region of the hf frequencies detection in HYSCORE experiments is usually limited by dead time of detection system to 20-30 MHz. Therefore, all hf interactions with higher frequencies presented in Table I are not visible in such experiment.

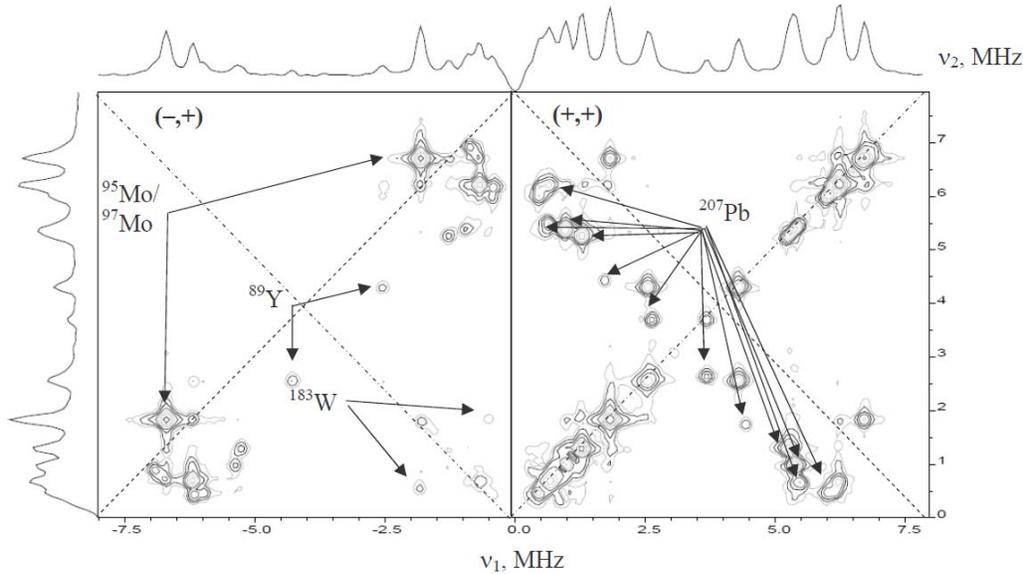

Fig. 5. The sum of X-band HYSCORE spectra measured in PbWO$_4$:Mo,Y crystal on the ED EPR position of $O_I$ center ($B_0$ = 343.12 mT) with different $\tau$ values. T = 18 K.

For the more precise determination of the shf parameters given in Table II, the angular dependence of HYSCORE spectra should be measured.

**C. Thermal stability and local energy levels of the hole centers**

The last column in Table I contains thermal stability parameters, such as the temperature of the thermal stability, ionization energy (trap depth), and frequency factor. These thermal parameters were roughly determined by the method of pulsed heating and in case of $O_I$ and $O_{II}$ centers they were more precisely determined by direct measurements of the temperature dependences of ionization probability of corresponding center.

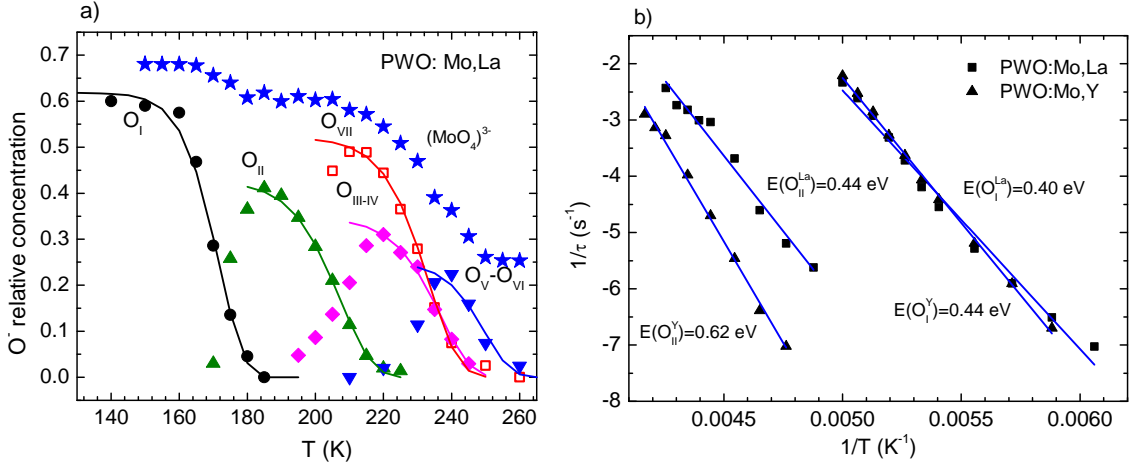

Fig. 6. (a) Dependence of $O^-$ and $(MoO_4)^{3-}$ ESR intensities in PWO:Mo,La on the temperature of pulsed annealing, showing the transformation of one hole center into another and recombination of freed holes with electrons stored at $(MoO_4)^{3-}$. The solid lines are the calculated ESR intensities of various $O^-$ centers performed using the first-order kinetics model with thermal parameters of hole traps presented in Table I. (b) The temperature dependence of the rate ionization of the $O_I$ and $O_{II}$ centers in PWO:Mo,La(Y). Solid lines are numerical fits to the data (see the text).

The thermal stability of $O^-$ centers was studied by the method of pulsed heating. After the UV irradiation at 140 K, the sample was heated up to a certain temperature $T_{an}$, hold at that temperature for three minutes, and then quickly cooled (with a rate of about 4 K/s) down to 90 K, where the ESR spectrum was measured. The 3 min interval was found as the best compromise to ensure a good thermalization of the sample and sufficient reproducibility of measured ESR intensities. The measurements at low temperature were necessary in order to avoid the influence of the spin-lattice relaxation on the spectral line width and signal intensity. The signal amplitudes obtained in such a way are depicted in Fig. 6. It can be seen that as a result of the heating, the spectrum of $O_I$ centers gradually vanishes and another spectrum appears, which also vanishes with the heating and other spectra appear, i.e. the hole centers are gradually transforming one into another. Obviously, such a transformation of $O^-$ centers may be due to their thermal ionization when part of the liberated holes can be re-captured at deeper traps.

In order to get a deeper insight into kinetics of the trapped holes and determine their thermal stability parameters, we have performed the numerical simulation of the ESR signal intensities. By neglecting the possibility of carriers recapture as suggested by the first-order kinetics, the concentration of localized holes is given by a simple exponential decay:

$$n = n_0 \exp(-P(T)t),$$

where $n_0$ is the initial concentration of trapped holes, $P(T)$ is the probability of their thermal ionization which is expressed in the form of Arrhenious law $P(T) = f_0 \exp(-E_t/kT)$, and $t$ is the annealing time. Here $f_0$ is a frequency factor and $E_t$ stands for the ionization energy (trap depth). The results of the simulation are shown in Fig. 6(a) by the solid lines for all the hole centers studied in PWO:Mo,La. For comparison, the thermal decay of the electrons trapped at Mo tetrahedra is shown as well. One can see that the thermal destruction of hole centers is accompanied with the corresponding decrease in the concentration of electron $(MoO_4)^{3-}$ centers. This proves the recombination of liberated holes with $(MoO_4)^{3-}$ centers which are stable up to higher temperatures. The determined thermal stability parameters (the ionization energies $E_t$ and frequency factors $f_0$) are listed in Table I.

In case of $O_I$ and $O_{II}$ centers, we were able to measure the decay rate $1/\tau(T) = P(T)$ of $O^-$ concentration after turning off the irradiation. The concentration decay of localized holes obeys simple exponential time dependence. Such dependence characterizes the first order kinetic process. The slope of logarithm of the ionization probability as a function of the inverse temperature gives the thermal ionization energy (see Fig. 6(b)) and the frequency factor $f_0 = N_c S v$ can be calculated as well.

In spite of the fact that g tensors of $O_{II}$ center in the both La and Y doped PWO crystals are practically the same, the thermal parameters, such as the thermal ionization energy, are distinctly different (see, e.g. Table I). This fact can be understood taking into account that g factors change only due to the spin-orbit coupling and therefore are not such sensitive to defect structure as a local energy level of a defect. The energy structure of the hole center can slightly vary depending on the position of perturbing impurities with respect to oxygen ions.

**D. TSL results and their correlation with ESR data**

The recombination process with participation of electron and hole centers can be studied by measurements of TSL. The TSL glow curves measured after the UV irradiation of the PWO:Mo,La #2 and PWO:Mo,Y crystals at various temperatures in the band-to-band (4.7 eV), exciton (around 4.1 eV), and Mo-related (around 3.8 eV) absorption regions are shown in Figs. 7 and 8. For comparison, the TSL curves of the PWO:Mo crystal as well as the undoped PWO crystal studied in [13, 24] are presented in Fig. 9. Figs. 7-9 illustrate that in all the crystals studied, the TSL glow curves are very complicated and consist of many strongly overlapping peaks.

The positions of complex TSL peaks and the intensity ratios of their separate components depend on the irradiation energy ($E_{irr}$), temperature ($T_{irr}$), and duration ($t_{irr}$). Therefore, by choosing the appropriate values of these parameters, it is possible to separate a TSL peak under investigation from the other peaks and to study it in more detail. The positions of the TSL peaks are shown in Table III.

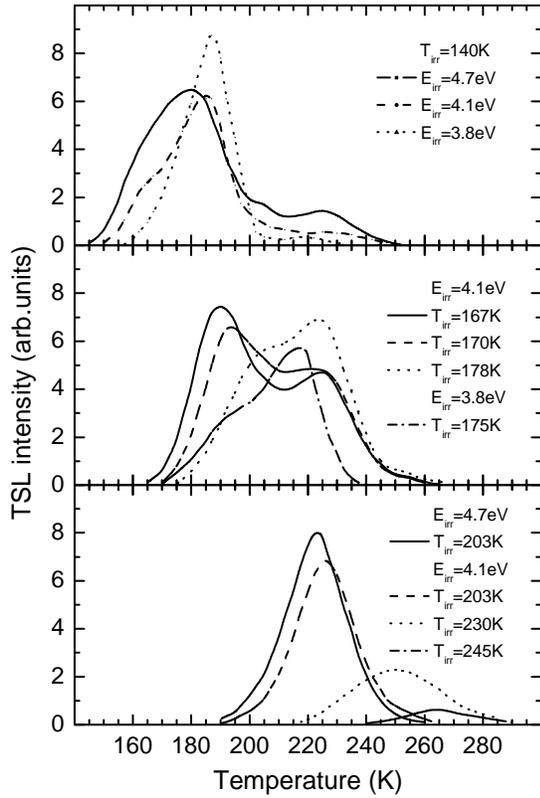

Fig. 7. TSL glow curves measured after irradiation of the PbWO$_4$:Mo,La #2 crystal for 15 minutes with different photon energies ($E_{irr}$) at different temperatures ($T_{irr}$).

The comparison of Figs. 7 and 8 with Fig. 9(a) indicates that the introduction of La$^{3+}$ and Y$^{3+}$ ions into a PWO:Mo crystal results in a strong change of its TSL characteristics (see also [15]). Indeed, after irradiation at 140-160 K, the most intense TSL peak of PWO:Mo is located around 222 K, while in PWO:Mo,La #2 and PWO:Mo,Y, the main TSL peak is located around 187 K and 192 K, respectively. Its complex structure reveals itself in the dependence of the peak position on $E_{irr}$ and $T_{irr}$ (see, e.g., Fig. 7 and Table 3). Weaker complex peaks are located at 218 K, 227 K in PWO:Mo,La #2 and at 226 K, 239 K in PWO:Mo,Y. In PWO:Mo,Y the main TSL peak is 2-2.5 times stronger than the 195 K peak in PWO:Mo, but the intensity of the 226 K or 238 K peak is much smaller as compared with the main TSL peak of PWO:Mo (35 times under $E_{irr}$=4.1 eV and about 300 times, under $E_{irr}$=3.8 eV).

The TSL glow curves of the PWO:Mo,La #1 crystal have been studied in [5, 12]. They are similar to those obtained for the PWO:Mo crystal (Fig. 9a), for the undoped PWO crystal grown by the Bridgman method [14], and for the undoped non-stoichiometric PWO crystals grown by the Czochralski method [16]. However, they strongly differ from the TSL glow curves of the other PWO:Mo crystals co-doped with trivalent rare-earth ions (Figs. 7 and 8; see also [4, 15]). For example, in the PWO:Mo,La #1 crystal, the total TSL intensity is by 2-3 orders of magnitude larger and the intensity of the TSL peak at 220-230 K is by 3 orders of magnitude larger than in the PWO:Mo,La #2 crystal. This effect is probably caused by accidentally non-stoichiometric composition of the former crystal. Indeed, a strong TSL and intense slow photostimulated G(II) emission, accompanying tunneling recombination processes [20], are observed in this crystal at T>160 K. This indicates to the presence of huge amount of various vacancy-related defects which causes a strong complexity of TSL characteristics of the PWO:Mo,La #1 crystal. Unlike this crystal, under photoexcitation of the PWO:Mo,La #2 and PWO:Mo,Y crystals, only the G(I) emission (2.37 eV, FWHM=0.47 eV at 80 K) is observed. The excitation band of this emission is located at 3.85 eV. The intensity of the G(I) emission in these crystals decreases twice at 175-180 K with the activation energy 0.23 eV.

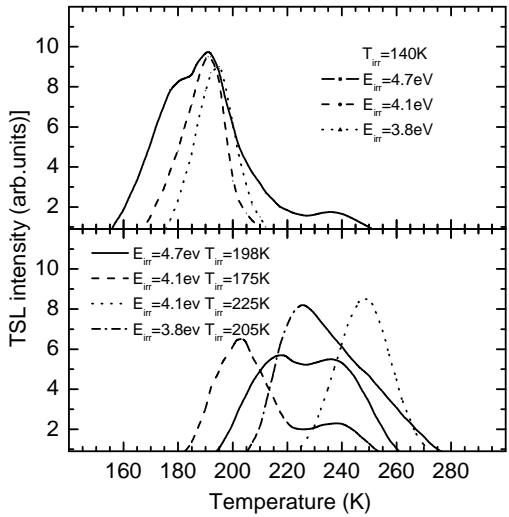

Fig. 8. TSL glow curves measured after irradiation of the PbWO$_4$:Mo,Y crystal for 15 minutes with different photon energies ($E_{irr}$) at different temperatures ($T_{irr}$).

As the paramagnetic O$^-$-type hole centers are not detectable in undoped PWO and in PWO:Mo, the TSL peaks, characteristic only for the La$^{3+}$- and Y$^{3+}$-codoped PWO:Mo crystals, are assumed to arise from these hole centers. To identify the origin of various TSL peaks, the parameters of the corresponding traps (trap depths $E_t$ and frequency factors $f_0$) were determined from the TSL data and compared with those obtained by the ESR method. For comparison, the $E_t$ values were calculated also for the PWO:Mo crystal. The $E_t$ values were determined by the partial cleaning method (for more details see [12]) from the ln $I_{TSL}$(1/T) dependences after heating of the UV-irradiated sample up to the selected temperature $T_{stop}$. Due to the complex structure of the TSL peaks, the $E_t$ values depend on $T_{stop}$. Therefore, they were obtained for many values of $T_{stop}$ (and also of $T_{irr}$) from the 150-240 K temperature range (Fig. 10). The deviations from the average $E_t$ values were 5-10 %. The frequency factors $f_0$ were calculated with the use

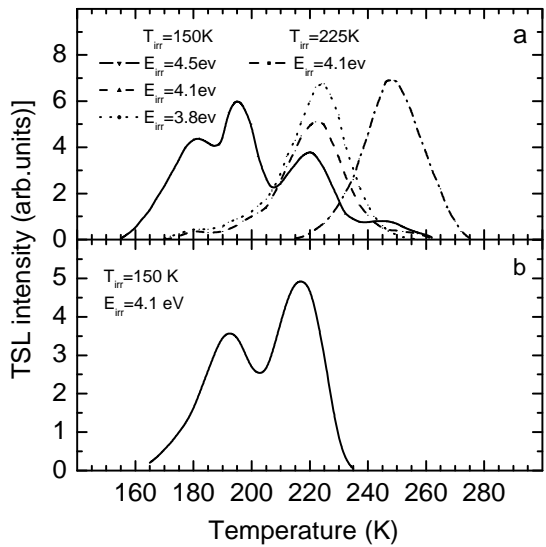

of the expression

$$f_0 = (\beta E_t / kT_m^2) \exp(E_t / kT_m),$$

where $\beta$ is the crystal heating rate, k is the Boltsman factor and $T_m$ is the maximum position of the TSL peak. The average values of the $E_t$ and $f_0$ parameters are shown in Table IV.

Fig. 9. TSL glow curves measured after irradiation of (a) the PbWO$_4$:Mo crystal and (b) the undoped PbWO$_4$ crystal for 15 minutes with different photon energies ($E_{irr}$) at different temperatures ($T_{irr}$).

For the 187 K and 227 K peaks in PWO:Mo,La #2

and 192 K and 238 K peaks in PWO:Mo,Y, the parameters are close to the $E_t$ and $f_0$ values obtained by the ESR method for hole $O_I$ and $O_{II}$ centers, respectively (Table 1). One can conclude that these TSL peaks are surely connected with the thermal destruction of $O_I$ and $O_{II}$ centers. For the PWO:Mo,La #1 crystal, smaller $E_t$ values (0.36 eV and 0.40 eV) are obtained for the TSL peaks observed at ≈186 K and ≈220 K, probably, due to their strong overlap with other peaks. For the weak TSL peaks around 250 K, the values of $E_t$ and $f_0$ in both PWO:Mo,La crystals are close to the parameters obtained by the ESR method for $O_{III}$, $O_{IV}$, $O_{VII}$ centers. However, these peaks are strongly overlapped with the 250 K peak which arises from the thermal destruction of electron $(MoO_4)^{3-}$ centers and has similar parameters. The values $E_t \approx 0.53$ eV, $f_0 \sim 10^9$ s$^{-1}$ are obtained for this peak in the PWO:Mo crystal where the O$^-$-type hole centers are not detectable in ESR. Note that for this crystal, much larger values of $E_t$ (0.60-0.75 eV) are obtained for the TSL peaks located around 190 K and in the 225-240 K temperature range as compared with those in PWO:Mo,La and PWO:Mo,Y (Fig. 10). The TSL peak at 265 K, probably arising from $O_V$, $O_{VI}$ centers, is also observed in both the PWO:Mo,La crystals studied (Fig. 7) but it is too weak to be analyzed. Note that due to different crystal heating conditions used at the TSL and ESR studies, each TSL peak is shifted towards higher temperatures with respect to the temperature ($T_q$) where the ESR signal intensity from the corresponding center decreases twice (see Table III).

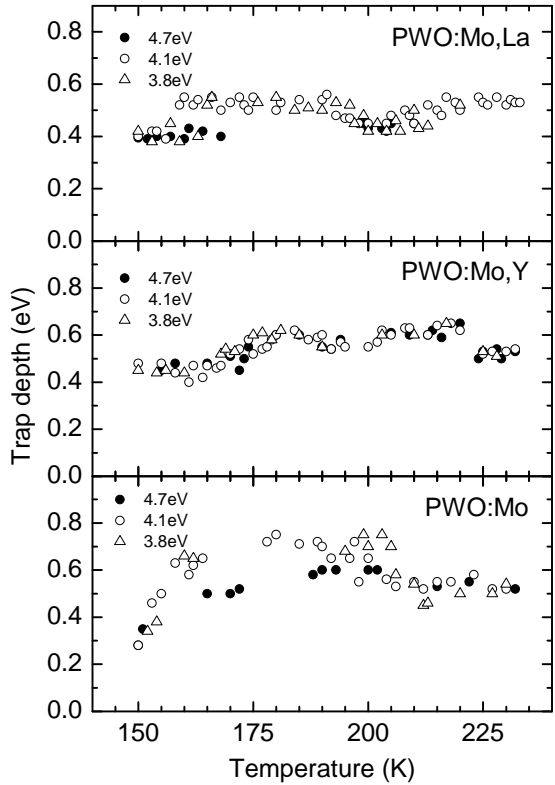

Fig. 10. Dependences of $E_t$ values on $T_{stop}$ or $T_{irr}$ measured for PWO:Mo,La, PWO:Mo,Y and PWO:Mo crystals under irradiation in the band-to-band (4.7 eV, closed circles), exciton (≈4.10 eV, open circles), and Mo-related (around 3.8 eV, open triangles) absorption regions.

As it is evident from Fig. 10, at the high-temperature side of the main TSL peak in PWO:Mo,La #2 and PWO:Mo,Y, an additional TSL peak is present whose parameters are similar to those obtained in [13] for electron Pb$^+$ centers located close to the oxygen vacancies of the type of WO$_3$: $E_t \approx 0.56$ eV, $f_0 \sim 10^{13}$ s$^{-1}$. In the undoped PbWO$_4$ crystals studied in [13, 24], this peak is observed at the same experimental conditions at about 193 K (Fig. 9(b)). One can conclude that the hole-related TSL peaks located at 187 K in PWO:Mo,La and at 192 K in PWO:Mo,Y are strongly overlapped with the ≈193 K TSL peak arising from $Pb^+$-$WO_3$ centers.

The TSL characteristics were studied also for the PWO:100 ppm Mo, 100 ppm Y crystal and the same peaks were observed at the TSL glow curve. However, the relative intensity of the ≈192 K peak in this crystal was about two times larger as compared with the PWO:1000 ppm Mo, 100 ppm crystal, probably, due to a larger concentration of oxygen vacancies. Like in the case of the PWO:1000 ppm Mo, 100 ppm Y crystal, four stages were clearly seen at the $E_t(T_{stop})$ dependence with the $E_t$ values of about 0.45 eV, 0.56 eV, 0.63 eV and 0.52 eV, arising from $O_I^-$, $Pb^+$-$WO_3$, $O_{II}^-$, and $(MoO_4)^{3-}$ centers, respectively (compare with Fig. 10).

The creation spectra of the TSL peaks in PWO:Mo,La #2 and PWO:Mo,Y are similar (Fig. 11). The peaks at 165-180 K, 227 K and 238 K, as well as the peak around 250 K (obtained after heating of the irradiated crystal up to 230 K followed by a quick cooling down to 140 K) are most effectively created under irradiation in the band-to-band and exciton absorption regions. The complex 187 K and 192 K peaks are effectively created also in the Mo-related region (see also Figs. 7 and 8). Similar creation spectrum was obtained in [4] for the TSL peak ascribed to $Pb^+$-$WO_3$ centers. This fact confirms our conclusion that the higher-temperature component of the main complex TSL peak of the crystals studied arises from the electron $Pb^+$-$WO_3$ centers.

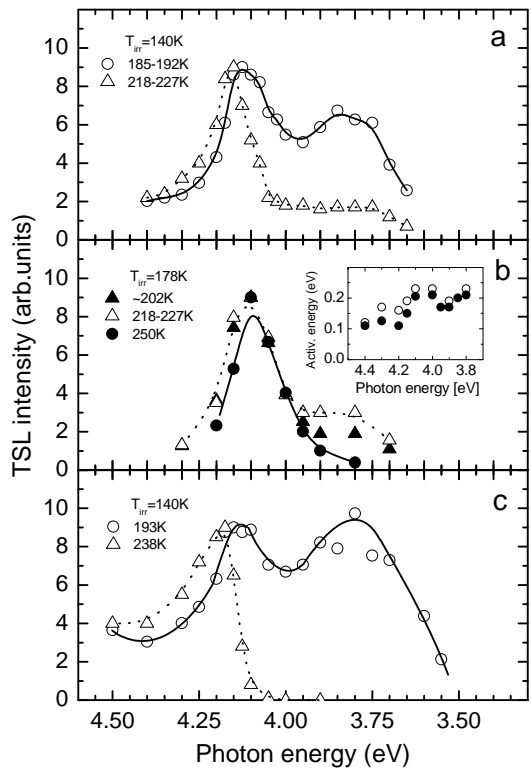

The activation energies $E_a$ for the creation of the TSL peaks at 187 K and 227 K in PWO:Mo, La #2 and at 192 K and 238 K in PWO:Mo,Y are similar and vary from 0.1 eV under irradiation at 4.4 eV to about 0.20 eV at $E_{irr}$<4.15 eV (see the inset in Fig. 11). The value of $E_a$=0.19 eV for the creation of hole $O_I$ centers in PWO:Mo,La was obtained also from the temperature dependence of the $O_I^{La}$ ESR intensity shown in Fig. 1.

Fig. 11. Creation spectra of some TSL peaks measured after the UV irradiation of the (a, b) PbWO$_4$:Mo,La #2 and (c) PWO:Mo,Y crystals at different temperatures ($T_{irr}$). In the inset: the dependence of the activation energy for the creation of the ≈187 K (closed circles) and 218-227 K (open circles) TSL peaks in the PWO:Mo, La #2 crystal on the irradiation energy $E_{irr}$.

At the initial irradiation stage, the linear dependence on the irradiation duration was obtained both for the number of paramagnetic $O_I$ centers and the TSL intensity around 190 K in the PWO:Mo, La #2 crystal.

In the TSL spectrum of undoped and Mo-doped PWO crystals, the G(II) emission (2.5 eV) was found to dominate [19-20]. The maximum of the TSL spectrum of the PWO:Mo,Y crystal, measured around 190 K and 235 K, is located at 2.4 eV (Fig. 12, solid line), i.e., closer to the Mo-related G(I) emission band which at any temperature in all the crystals studied is peaking at 2.37 eV (dotted line). This allows to suggest that in the mentioned crystals, thermally released holes recombine mainly with electrons stored at $(MoO_4)^{3-}$ centers. This recombination process is directly visible in temperature decay of $(MoO_4)^{3-}$ concentration shown in Fig. 6(a). After heating of the irradiated PWO:Mo crystal up to 230 K, mainly the Mo-related 250 K peak remains at the glow curve. The TSL spectrum in this temperature range has the maximum at 2.48 eV (dashed line). One can conclude that the recombination of thermally released electrons with the trapped holes results in the appearance of the G(II) emission. The dominating complex TSL peak, observed in these crystals around 222 K, can arise from the thermal destruction of both the non-paramagnetic electron centers (e.g., clusters of oxygen vacancies with two trapped electrons) and non-paramagnetic hole centers (e.g., lead vacancies with two trapped holes). Indeed, the noticeable reduction of the number of electron $(MoO_4)^{3-}$ and $(CrO_4)^{3-}$ centers in PWO has been observed just in this temperature range [12, 14], although the thermally stimulated release of electrons from these centers takes place at higher temperatures. This effect can be caused by the thermal destruction of the above mentioned non-paramagnetic hole centers and recombination of the released holes with $(MoO_4)^{3-}$ or $(CrO_4)^{3-}$ groups (see also [24]). If so, the complex band consisting of both the G(II) and G(I) emission should be observed in the TSL spectrum around 222 K. Indeed, a wide (FWHM=0.6 eV) emission peaking at 2.45 eV is stimulated in this temperature range.

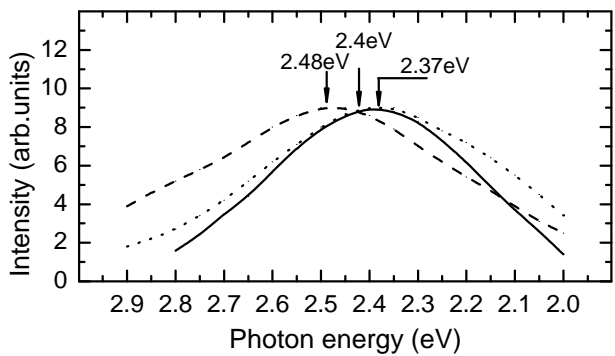

Fig. 12. TSL spectra measured at 230-240 K for the PWO:Mo,Y crystal (solid line) and around 250 K for the PWO:Mo crystal (dashed line). The G(I) emission spectrum measured at 240 K under 3.6 eV excitation (dotted line).

### IV. CONCLUSIONS

Detailed ESR measurements of $PbWO_4$:Mo,La(Y) single crystals have shown that the holes created by the UV irradiation can be trapped at regular oxygen ions nearby a perturbing defect such as a lead vacancy, an impurity ion (La, Y, Mo) or another unidentified lattice perturbation, forming a variety of O$^-$ centers with the thermal stability from about 170 K up to 240 K. The hole state (an electron deficiency at an oxygen ion) is essentially delocalized over the surrounding ions resulting in rich hf structure of corresponding ESR spectrum. The same ESR characteristics have been observed for the samples of different origin and purity. Therefore, the behavior of the hole centers described above seems to be characteristic of PbWO$_4$:Mo,La(Y) itself. However, we emphasize that described here O$^-$ centers were observed only in the doubly (Mo,La and Mo,Y) doped crystals. In our opinion, the presence of Mo impurity was necessary in order to create a measurable number of traps for electrons and consequently increased number of trapped holes. La$^{3+}$ and Y$^{3+}$ impurities also enable partial charge compensation of lead vacancies in the vicinity of which the single hole can be localized at oxygen ion to provide the full charge compensation of the lead vacancy defect. Without these impurities, holes should be preferably trapped by a single lead vacancy forming the non paramagnetic {V$_{Pb}$+2h}-type centers.

The thermal release of the trapped holes and their subsequent radiative recombination with the electrons stored at different traps, including Pb$^+$-WO$_3$ and (MoO$_4$)$^{3-}$ centers, results in the appearance of the TSL peaks located around 190 K and in the 225-250 K temperature range. The comparison of the thermal stability parameters obtained by the ESR method for hole O$_I$ and O$_{II}$ centers with those obtained for the TSL peaks at 187 K and 227 K in PWO:Mo,La #2 and at 192 K and 238 K in PWO:Mo,Y allows the conclusion that the above-mentioned TSL peaks are surely connected with the thermal destruction of the O$_I$ and O$_{II}$ centers. These peaks are most effectively created in the band-to-band and exciton absorption region, i.e. at E$_{irr}$>4 eV. The activation energy for the hole centers creation under excitation in the exciton region is about 0.2 eV.

## ACKNOWLEDGEMENTS

The authors gratefully acknowledge the financial support of the Czech project GA AV IAA100100810, the SAFMAT project CZ.2.16/3.1.00/22132, and the Estonian Science Foundation project No. 8678. The authors are also grateful to Prof. Pöppl (University of Leipzig) for the useful discussion of the results obtained.

Table I. Spectral characteristics of hole traps in PbWO$_4$:Mo,La(Y) crystals obtained from the ESR data. Principal axis directions of g tensors are given by polar (θ) and azimuthal (φ) angles relative to *a,b,c* crystal axes and presented for one of four equivalent centers. The error margin of the polar and azimuthal angles is approximately 2 deg.

| Center | g tensor | Principal axes | | HF interaction[*] | Thermal stability |
| --- | --- | --- | --- | --- | --- |
| | | θ | φ | ($10^{-4}$ cm$^{-1}$) | Parameters |
| $O_I^{La}$ | $g_{[001]}$: 2.027(5) | not | | $A_{La}$: 0.9 mT | ~165 K [**] |
| | $g_{[100]}$: 2.010(5) | determ. | | | $E_t$=0.40(2) eV, $f_0$≈10$^9$ 1/s |
| $O_I^Y$ | $g_1$: 2.0142(5) | 46 | 351 | $A^W$: 1.5 mT | ~160 K |
| | $g_2$: 2.0274(5) | 78 | 93 | $A^{Pb}$: 2.8 mT | $E_t$=0.44(2) eV, $f_0$≈10$^{10}$ 1/s |
| | $g_3$: 2.0503(5) | 47 | 194 | | |
| $O_{II}$ | $g_1$: 1.9973(5) | 100 | 22 | $A_1^{Pb}$: 34.8 mT | ~190-210 K |
| | $g_2$: 2.0094(5) | 134 | 122 | $A_2^{Pb}$: 1.4 mT | $E_t^{La}$ = 0.44(2) eV, $f_0^{La}$ ≈ 10$^8$ 1/s |
| | $g_3$: 2.0400(5) | 45 | 102 | $A^W$: 1.4 mT | $E_t^Y$ = 0.62(2) eV, $f_0^Y$ ≈ 10$^{12}$ 1/s |
| $O_{III}$ | $g_1$: 1.9997(2) | 91 | 350 | not resolved | ~220 K |
| | $g_2$: 2.0089(2) | 127 | 80 | | $E_t$=0.56(4) eV, $f_0$ ~10$^9$ 1/s |
| | $g_3$: 2.0350(2) | 37 | 79 | | |
| $O_{IV}$ | $g_1$: 2.0027(2) | 106 | 29 | $A^W$: 5.6 mT | ~220 K |
| | $g_2$: 2.0064(2) | 41 | 319 | $A_1^{Pb}$: 1.9 mT | $E_t$=0.56(4) eV, $f_0$ ~10$^9$ 1/s |
| | $g_3$: 2.0128(2) | 53 | 106 | $A_2^{Pb}$: 0.5 mT | |
| $O_{V-VI}$ | $g_{[001]}$: 2.035(1) | not | | not resolved | ~240 K |
| | $g_{[100]}$: 2.015(1) | determ. | | | $E_t$=0.72(4) eV, $f_0$ ~10$^{12}$ 1/s |
| $O_{VII}$ | $g_1$: 2.0001(4) | 102 | 119 | $A^W$: 4.58 mT | ~210 K |
| | $g_2$: 2.0007(4) | 43 | 197 | $A_1^{Pb}$: 1.15 mT | $E_t$=0.55(4) eV, $f_0$ ~10$^9$ 1/s |
| | $g_3$: 2.0029(4) | 49 | 40 | $A_2^{Pb}$: 0.52 mT | |

[*] HF splitting is measured along *c* axis.

[**] Temperature of the thermal stability, see Fig. 6.

Table II. The hf parameters of O$_I$ center describing interaction of the hole with surrounding nuclei obtained by HYSCORE spectra analysis for the orientation (**B** ΙΙ [001]).

| Nucleus | Cross peaks (MHz) | $|A|$ (MHz) | $|B|$ (MHz) |
|---|---|---|---|
| $^{95,97}$Mo | [-6.71, 1.82]<br>[-1.82, 6.71] | 8.26 | ~0.12 |
| $^{89}$Y | [-4.30, 2.54]<br>[-2.54, 4.3] | 6.84 | ~0.09 |
| $^{181}$W | [-1.82, 0.52]<br>[-0.52, 1.82] | 2.34 | ~0.03 |
| $^{207}$Pb(1) | [0.65, 6.12]<br>[6.12, 0.52] | 5.60 | ~0.22 |
| $^{207}$Pb(2) | [0.65, 5.47]<br>[5.47, 0.65] | 4.82 | - |
| $^{207}$Pb(3) | [0.98, 5.34]<br>[5.34, 0.98] | 4.36 | ~0.11 |
| $^{207}$Pb(4) | [1.30, 5.27]<br>[5.27, 1.30] | 3.97 | ~0.21 |
| $^{207}$Pb(5) | [1.69, 4.43]<br>[4.43, 1.69] | 2.73 | - |
| $^{207}$Pb(6) | [2.60, 3.65]<br>[3.65, 2.60] | 1.04 | ~0.11 |

Table III. Positions of some TSL glow peaks (K) obtained after irradiation of the PWO:Mo,La #2, PWO:Mo,Y, and PWO:Mo crystals in the band-to-band (4.7 eV), exciton (around 4.1 eV), and Mo-related (around 3.8 eV) absorption regions. The most intense peaks are shown in bold.

| Irrad. energy | Temperature of TSL peaks (K) | | |
|---|---|---|---|
| | PWO:Mo,La #2 | PWO:Mo,Y | PWO:Mo |
| 4.7 eV | **180** 202 224 250 | **180** **191** 203 217 237 250 | 182 195 **220** 248 |
| 4.1 eV | **185-190** 202 227 250 | **180** **192** 203 238 249 | 182 200 **222** 249 |
| 3.8 eV | **187** 202 218 250 | **194** 226 250 | 181 197 **224** 249 |

Table IV. Trap depths ($E_t$) and frequency factors ($f_0$) for the hole-related TSL peaks in the PbWO$_4$:Mo,La #2 and PWO:Mo,Y crystals.

| Center | Temp. of EPR signal decay (K)[*)] | Temperature of TSL peak (K) | Trap depth $E_t$ (eV) | Frequency factor $f_0$ (1/s) |
|---|---|---|---|---|
| $O_I^{La}$ | 170 | 187 | 0.39(2) | ~$10^9$ |
| $O_{II}^{La}$ | 210 | 227 | 0.45(2) | ~$10^8$ |
| $O_{III,IV,VII}$ | 230-235 | 247, 251 | 0.54(2) | ~$10^9$ |
| $O_{V,VI}$ | 250 | 265 | not deter. | not deter. |
| $O_I^Y$ | 180 | 192 | 0.45(2) | ~$10^{10}$ |
| $O_{II}^Y$ | 220 | 238 | 0.63(2) | ~$10^{12}$ |

[*)] Temperature where the EPR signal intensity from the corresponding center decreases about twice, see Fig. 6(a)